\documentclass{PoS}

\usepackage[numbers]{natbib}

\newcommand{\msun}{\ensuremath{M_\odot}}
\newcommand{\Chimera}{\textsc{Chimera}}
\newcommand{\figwidth}{.7\columnwidth}
\newcommand{\Rshock}{\ensuremath{R_{\rm sh}}}

\title{Two- and three-dimensional simulations of core-collapse supernovae with CHIMERA}

\ShortTitle{Two- and three-dimensional simulations of core-collapse supernovae with CHIMERA}

\author{\speaker{Eric J. Lentz}$^{ab}$, Stephen W. Bruenn$^{c}$, J.~Austin Harris$^{a}$, Merek Austin Chertkow$^{a}$, W.~Raphael Hix$^{ba}$,    Anthony Mezzacappa$^{ba}$, O.~E.~Bronson Messer$^{da}$, John~M.~Blondin$^{e}$,  Pedro Marronetti$^{c}$, Christopher~M. Mauney$^{e}$, and Konstantin~N.~Yakunin$^{c}$\thanks{
This research was supported by the U.S. Department of Energy Offices of Nuclear Physics and Advanced Scientific Computing Research;
 the NASA Astrophysics Theory and Fundamental Physics Program (grants NNH08AH71I and NNH11AQ72I); and the National Science Foundation PetaApps Program (grants  OCI-0749242, OCI-0749204, and OCI-0749248).
This research was also supported by
the NSF  through TeraGrid resources provided by the National Institute for Computational Sciences under grant number TG-MCA08X010; 
resources of the National Energy Research Scientific Computing Center, supported by the U.S. DoE Office of Science under Contract No. DE-AC02-05CH11231; and
an award of computer time from the Innovative and Novel Computational Impact on Theory and Experiment (INCITE) program at the Oak Ridge Leadership Computing Facility, supported by the  U.S. DoE Office of Science under Contract No. DE-AC05-00OR22725.
}\\
\llap{$^a$}Department of Physics \& Astronomy, University of Tennessee\\ 
   Knoxville, TN 37996, USA\\
\llap{$^b$}Physics Division, Oak Ridge National Laboratory\\
   Oak Ridge, TN 37831, USA\\
\llap{$^c$}Department of Physics \& Astronomy, Florida Atlantic University\\ 
   Boca Raton, FL 33431, USA\\
\llap{$^d$}National Center for Computational Sciences, Oak Ridge National Laboratory\\ 
   Oak Ridge, TN 37831, USA\\
\llap{$^e$} Department of Physics, North Carolina State University\\
   Raleigh, NC 27695, USA \\

           E-mail: \email{elentz@utk.edu}}

\abstract{Ascertaining the core-collapse supernova mechanism is a complex, and yet unsolved, problem dependent on the interaction of general relativity, hydrodynamics, neutrino transport, neutrino-matter interactions, and nuclear equations of state and reaction kinetics. \emph{Ab initio} modeling of core-collapse supernovae and their nucleosynthetic outcomes requires care in the coupling and approximations of the physical components. We have built our multi-physics \Chimera\ code for supernova modeling in 1-, 2-, and 3-D, using ray-by-ray neutrino transport, approximate general relativity, and detailed neutrino and nuclear physics. We discuss some early results from our current series of  exploding 2D simulations and our work to perform computationally tractable simulations in 3D using the ``Yin--Yang'' grid.}

\FullConference{XII International Symposium on Nuclei in the Cosmos\\
		 August 5-12, 2012\\
		 Cairns, Australia}

\begin{document}

\section{Simulating the revival of the supernova shock}

Reviving the stalled shock above the nascent neutron star in core-collapse supernovae (CCSNe) has been a vexing and compelling problem in computational astrophysics for more than 40 years.
The cumulative analysis of CCSN simulations has shown that there are many physical ingredients that impact the shock revival, some enhancing revival and others suppressing it \cite{Mezz05,Jank12,KoTaSu12}.
Important microphysical inputs include the various neutrino-matter interactions and the nuclear equation of state (EoS), while macroscopic inputs include the treatment of hydrodynamics, general relativity, and neutrino transport.

Shock revival is subject to physical instabilities that make CCSNe inherently an asymmetric problem.
Neutrino-driven convection not only induces non-spherical motion, but more critically boosts the rate of energy deposition in the `hot mantle' between the proto-neutron star (PNS) and the shock.
The shock's  instability against non-radial perturbations,  the Standing Accretion Shock Instability (SASI),  is dominated in 2D (axisymmetry) by an $\ell = 1$ `sloshing' mode that expands the volume inside the shock.
The flow onto the PNS is channelled into (intermittent) accretion streams.
These features make the dynamics of CCSNe intrinsically non-spherical.
Here we report on some aspects of our current series of 2D simulations, which show shock expansion to many thousand km and `diagnostic' explosion energies \cite{SuKoTa10} close to the canonical 1~B ($\equiv10^{51}$~erg) \cite{BrMeHi13} and look ahead to our forthcoming 3D simulation.
The 2D simulations begin from the 12, 15, 20, and 25~\msun\ progenitors of \citet{WoHe07} with 512 adaptive radial zones and 256 angular zones covering the entire sphere in our \Chimera\ code \cite{BrMeHi13}.
The fluid motion is solved using the VH1 implementation of the dimensionally split, Lagrangian-remap (PPMLR) scheme, with a multipole  expansion of gravity that includes a GR monopole term.
The neutrino transport is a ``ray-by-ray-plus'' (RbR) scheme using GR-enhanced, multi-group flux-limited diffusion \cite{Brue85} with the best available neutrino-matter interactions.
We utilize the $K=220$~MeV version of the Lattimer-Swesty \cite{LaSw91} EoS for $\rho > 10^{11}$~g~cm$^{-3}$ and Cooperstein \cite{Coop85} EoS for NSE where $\rho < 10^{11}$~g~cm$^{-3}$ and an extended version of the Cooperstein $e^-e^+\gamma$-EoS throughout.
Outside NSE conditions we use the integrated 14-species XNet \cite{HiTh99b} $\alpha$-network.

\begin{figure}[h]
\center
\includegraphics[width=\figwidth]{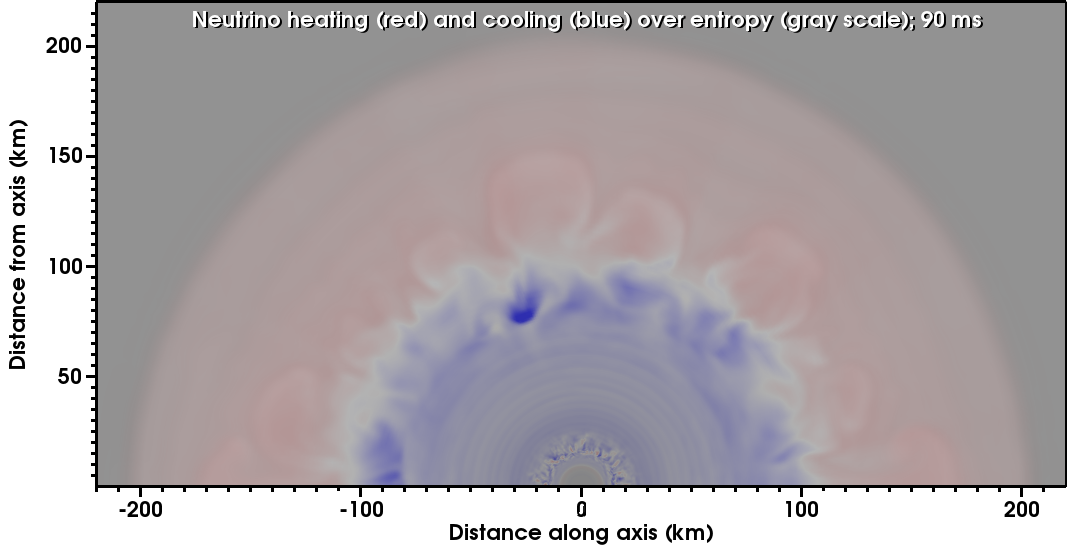}
\includegraphics[width=\figwidth]{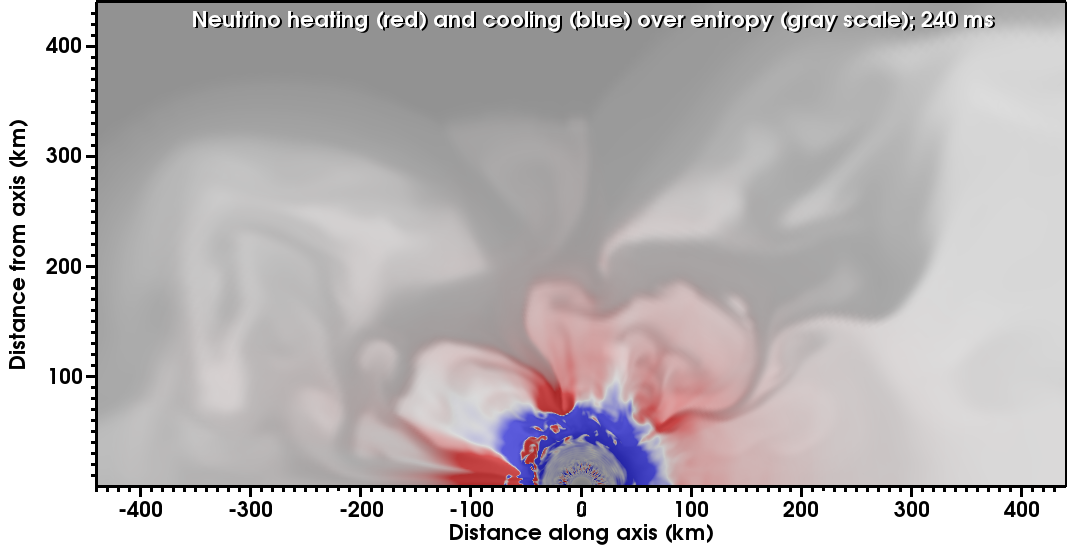}
\caption{Net neutrino heating (red) and cooling (blue) on a grayscale entropy background for 12 \msun\ model at 90~ms (top) and 240~ms (bottom) after bounce. The transition from heating to cooling is known as the gain `radius'.  Darker grays indicate lower entropy regions. \label{fig:heat}}
\end{figure}

\section{Dynamics during shock revival}

After a brief epoch ($\sim$10--40~ms after bounce) of prompt convection of lepton and entropy unstable gradients left in the wake of the rebound shock, the hot mantle becomes relatively quiescent until neutrino heating results in convective instability. By 90~ms after bounce in our 12-\msun\ model (Fig.~\ref{fig:heat}; top) $\sim$10 convective plumes are rising from the heating/cooling interface (gain radius), but have yet to distort the spherical shape of the shock.
As the plumes grow in size they reduce in number  and help trigger growth of the SASI sloshing mode.
These instabilities help to form a configuration common to 2D CCSN simulations \cite{BuJaRa06,BuLiDe07,MaJa09,MuJaMa12} with the shock elongated along the axis, large plumes rising near the poles, and an accretion stream forming at the shock cusp between lobes where the shock radius is typically at a minimum.
A different configuration found in some 2D CCSN simulations can occur if the accretion stream forms at, or near, one pole resulting in a single lobed configuration. 
This configuration does not appear in this series of models, but it did occur in the 20~\msun\ model in our previous series \cite{BrMeHi09b} and some models of  \cite{BuLiDe06,SuKoTa10,MuJaMa12}.

\begin{figure}[h]
\center
\includegraphics[width=\figwidth]{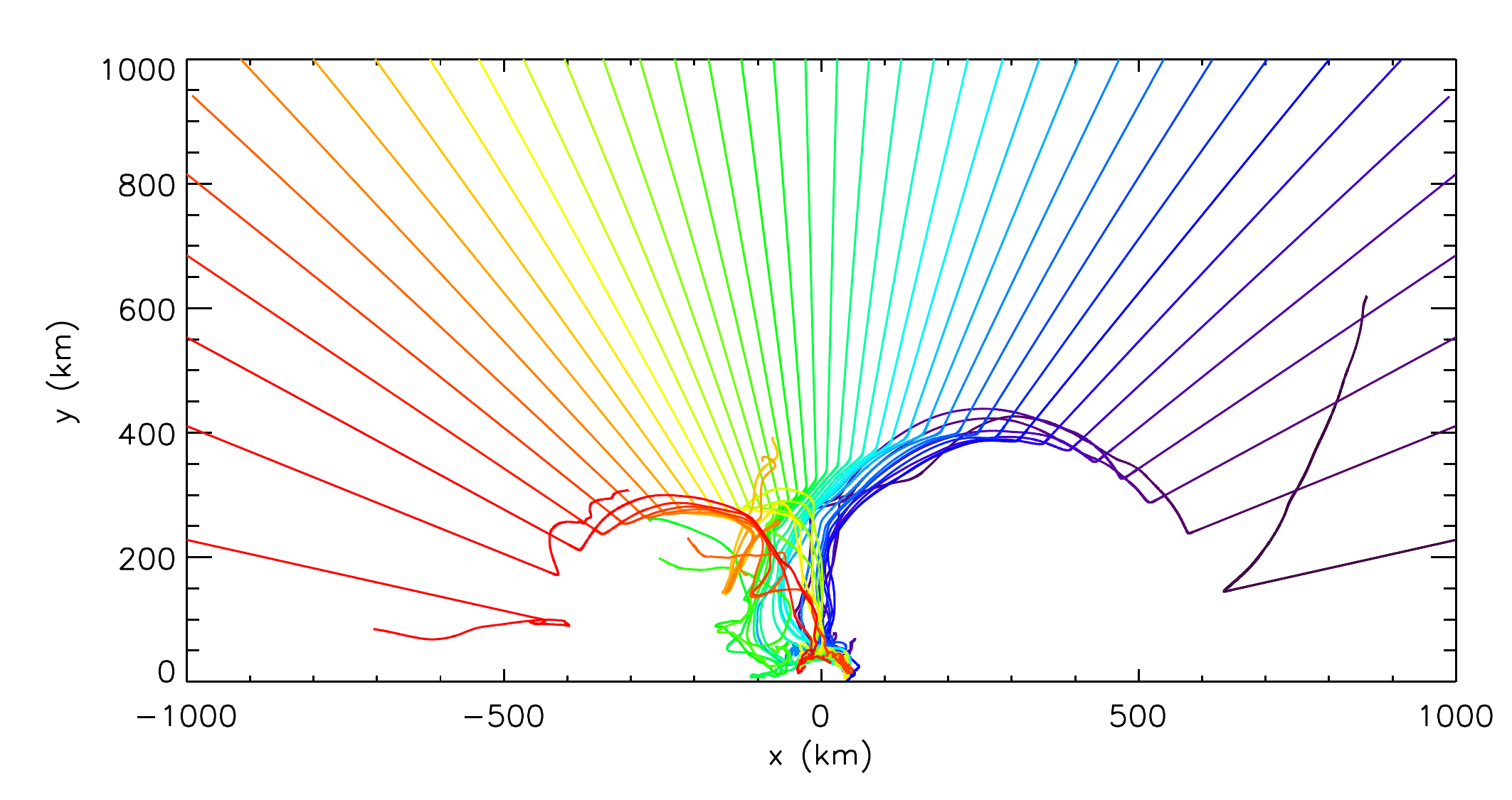}
\caption{Trajectories for a rows of tracers initialized at $M=1.393\,\msun$ in the Si-shell in the 12 \msun\ model colored by starting angle. \label{fig:tracer}}
\end{figure}

At 240~ms (Fig.~\ref{fig:heat}; bottom) the shock has a minimum  $\Rshock\sim$350~km near the equator with a low entropy stream (darker shades).
In the inner 200~km, there are 2~major streams reaching the PNS near the equator and on the left  and a thinner stream on the right. Where these penetrate to the PNS there is a localized, intense heating as shown by the deep red color. Because of the RbR transport, the strongest heating is only above the `hot spots' where the streams decelerate at the PNS.
We would expect to find some lateral spreading of these intense heating regions in simulations with full multi-D transport (i.e., without the RbR-approximation) \cite{OtBuDe08}. Along the right ($+$) pole where there is no active accretion, there is less intensive and smoother heating in a polar outflow.

A useful approach to diagnosing fluid motion is with `Lagrangian tracer particles'.
These `tracers' are initialized  in equal-mass spaced  `rows' of 40 tracers with an equal volume lateral spacing within a row \cite{ChMeHi12}.
Fig.~\ref{fig:tracer} shows the trajectories of a row of tracers that start from inside the O-rich Si-layer of the progenitor.
This row of tracers strikes the shock when $\Rshock\sim$500~km  and the shock is already significantly distorted and beginning to revive.
The tracer closest to each pole in this row is deflected into a lobe of the expanding shock of the nascent CCSN explosion.
Most of the remaining tracers are deflected along the inward-curving shock into the equatorial accretion stream and plunge toward the PNS.
Tracer analysis provides an alternative view of the supernova dynamics than the commonly used `snapshot' images.
We end the plotting of these traces at the time when most of the tracers reach the PNS to illustrate in the infall stage of their motion, but subsequent tracer analyses will help identify how material fills the plumes and becomes ejecta.

\begin{figure}[h]
\center
\includegraphics[width=\figwidth]{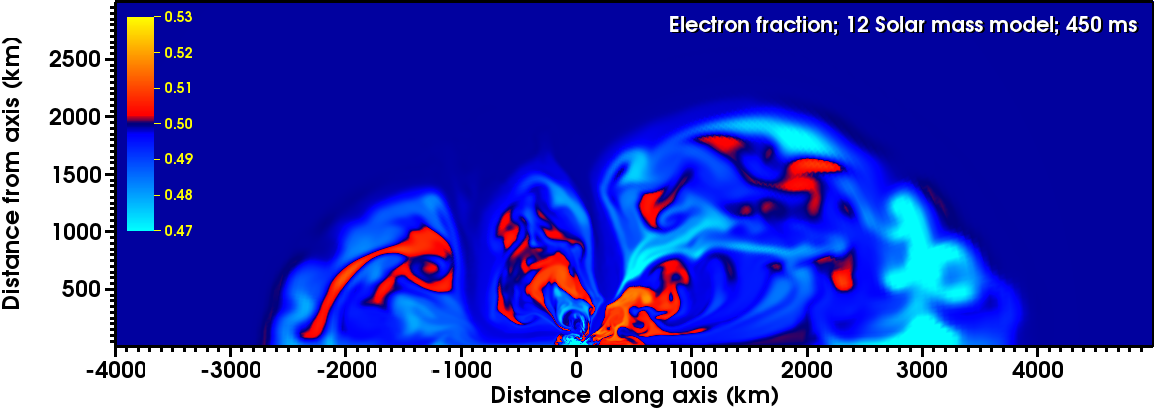}
\includegraphics[width=\figwidth]{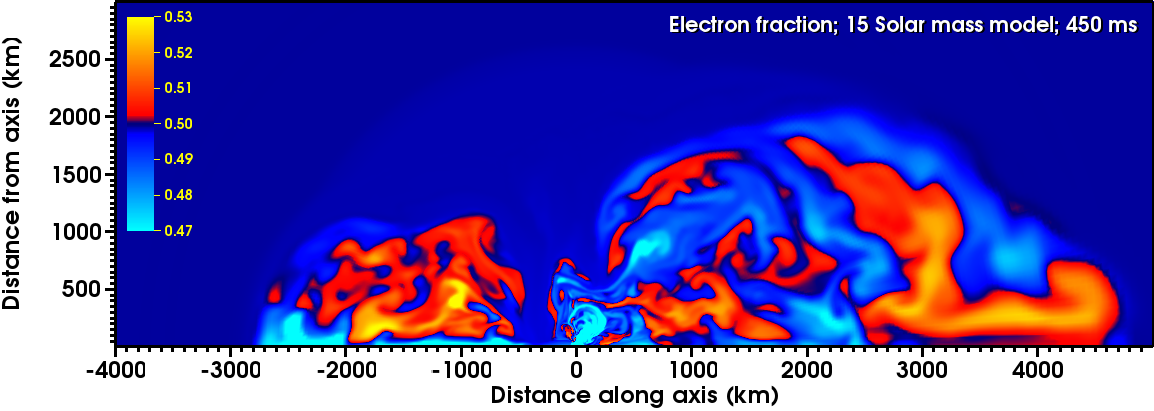}
\includegraphics[width=\figwidth]{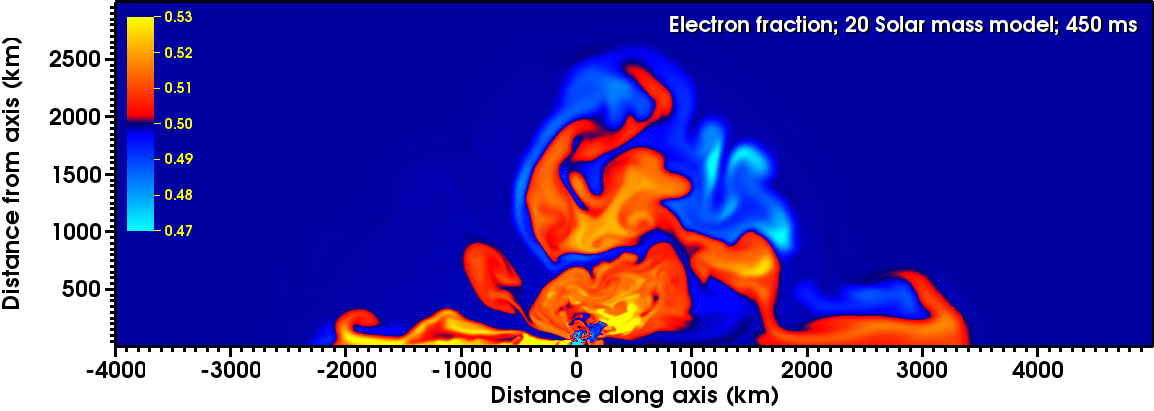}
\includegraphics[width=\figwidth]{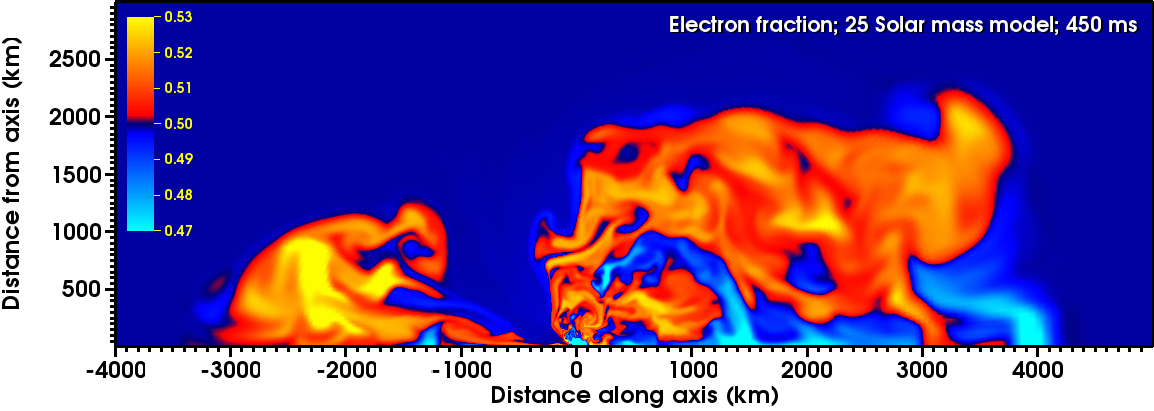}
\caption{Electron fraction, $Y_e$, at 450~ms after bounce for all models. \label{fig:Ye}}
\end{figure}

\section{Ejecta}

Though these models are not complete enough for detailed nucleosynthesis (the 12~\msun\ model at 850~ms after bounce with an asymptotic explosion energy and negligible accretion seems to be nearing that point \cite{BrMeHi13}), by examining the electron fraction, $Y_e$, of the ejecta at 450~ms for all models (Fig.~\ref{fig:Ye}) we can see that proton-richness of the ejecta appears to increase with progenitor mass.
For the 12~\msun\ model, there is more lower-$Y_e$ (lighter blues) moderately neutron-rich ejecta than higher-$Y_e$, moderately proton-rich ejecta (red--yellow), and $Y_e <  0.52$.
For the 25~\msun\ model there is a shift to higher-$Y_e$ ejecta with notable fractions above $Y_e=0.52$, though there remains some low-$Y_e$ material visible.
The 15 and 20~\msun\ models are clearly intermediate cases, but it is not clear at this point if there is a distinct sequence or variation around a trend.
Comparison of the $Y_e$ maps to the velocity maps show that the vast majority of material deviating from the $Y_e \approx 0.5$ background is moving outward, and for the case of the 12~\msun\ model remains so after an additional 400~ms of evolution.
For all models we find both lower- and higher-$Y_e$ ejecta with complex, intermixed distributions.
Because detailed nucleosynthesis must await the completion of the explosive phase before a definitive measure of the composition of the ejecta can be made using our tracers, we refrain from any premature quantification of the ejecta  $Y_e$ except to highlight the potential for considerable differences in ejecta composition across progenitor mass.

\section{Summary and a look ahead}

Our four axisymmetric models all show large scale shock expansion to many thousand km with `diagnostic' energies of significant fraction of the canonical 1~Bethe($\equiv10^{51}$~erg) explosion energy \cite{BrMeHi13}. Though the 12~\msun\ model at 850~ms may have completed the accretion/neutrino heating phase, the other models at only 500~ms are clearly incomplete.
From the models we can view a number of primary and secondary quantities to better understand how these explosions occur.
The pattern of neutrino heating and cooling can show  the source of initial neutrino-driven convection and the intense heating at the bottom of the accretion funnels formed by the curvature of the aspherical shock.
The flow of the accreted fluid into these funnels can also be clearly seen by following our tracer particles, which will eventually be used to compute detailed nucleosynthesis of the ejecta in our finalized simulations.

The imposition of axisymmetry can affect the development of CCSNe and to understand these effects requires models without imposed dimensional symmetries.
We are working to integrate the `Yin--Yang' grid \cite{KaSa04} into \Chimera. The `Yin--Yang' grid gains advantage by eliminating the constricted polar zones of standard spherical-polar grids and the time-step limitations that they bring.
During our testing phase, we constructed a low-resolution ($\approx$3.3$^{\circ}$) pilot run, initiated from a perturbed spherically symmetric model $\approx$50~ms after bounce. This pilot run, over the course of 100~ms evolution, shows neutrino driven convection as well as hints of an early spiral SASI mode.
The shock position (viewed in a slice through the center) appears to rotate like a disk with an off-center pivot on to which is superimposed  the distortions of of the shock surface from rising convective plumes.
We have recently initiated a better resolved ($\approx$1.3$^{\circ}$) run using the same complete physics set of our 2D simulations.
We will present details of this model in  future papers.

\bibliographystyle{apsrev}
\bibliography{mn-jour,add_journals,supernova,network,hydro-mhd}

\begin{thebibliography}{19}
\expandafter\ifx\csname natexlab\endcsname\relax\def\natexlab#1{#1}\fi
\expandafter\ifx\csname bibnamefont\endcsname\relax
  \def\bibnamefont#1{#1}\fi
\expandafter\ifx\csname bibfnamefont\endcsname\relax
  \def\bibfnamefont#1{#1}\fi
\expandafter\ifx\csname citenamefont\endcsname\relax
  \def\citenamefont#1{#1}\fi
\expandafter\ifx\csname url\endcsname\relax
  \def\url#1{\texttt{#1}}\fi
\expandafter\ifx\csname urlprefix\endcsname\relax\def\urlprefix{URL }\fi
\providecommand{\bibinfo}[2]{#2}
\providecommand{\eprint}[2][]{\url{#2}}

\bibitem[{\citenamefont{{Mezzacappa}}(2005)}]{Mezz05}
\bibinfo{author}{\bibfnamefont{A.}~\bibnamefont{{Mezzacappa}}},
  \bibinfo{journal}{Annu. Rev. Nucl. Part. Sci.} \textbf{\bibinfo{volume}{55}},
  \bibinfo{pages}{467} (\bibinfo{year}{2005}).

\bibitem[{\citenamefont{{Janka}}(2012)}]{Jank12}
\bibinfo{author}{\bibfnamefont{H.-T.} \bibnamefont{{Janka}}},
  \bibinfo{journal}{Annu. Rev. Nucl. Part. Sci.} \textbf{\bibinfo{volume}{62}},
  \bibinfo{pages}{407} (\bibinfo{year}{2012}), \eprint{1206.2503}.

\bibitem[{\citenamefont{{Kotake} et~al.}(2012)\citenamefont{{Kotake},
  {Takiwaki}, {Suwa}, {Iwakami Nakano}, {Kawagoe}, {Masada}, and
  {Fujimoto}}}]{KoTaSu12}
\bibinfo{author}{\bibfnamefont{K.}~\bibnamefont{{Kotake}}},
  \bibinfo{author}{\bibfnamefont{T.}~\bibnamefont{{Takiwaki}}},
  \bibinfo{author}{\bibfnamefont{Y.}~\bibnamefont{{Suwa}}},
  \bibinfo{author}{\bibfnamefont{W.}~\bibnamefont{{Iwakami Nakano}}},
  \bibinfo{author}{\bibfnamefont{S.}~\bibnamefont{{Kawagoe}}},
  \bibinfo{author}{\bibfnamefont{Y.}~\bibnamefont{{Masada}}}, \bibnamefont{and}
  \bibinfo{author}{\bibfnamefont{S.-i.} \bibnamefont{{Fujimoto}}},
  \bibinfo{journal}{Advances in Astronomy} \textbf{\bibinfo{volume}{2012}},
  \bibinfo{eid}{428757} (\bibinfo{year}{2012}).

\bibitem[{\citenamefont{{Suwa} et~al.}(2010)\citenamefont{{Suwa}, {Takiwaki},
  {Whitehouse}, {Liebend{\"o}rfer}, and {Sato}}}]{SuKoTa10}
\bibinfo{author}{\bibfnamefont{Y.}~\bibnamefont{{Suwa}}},
  \bibinfo{author}{\bibfnamefont{T.}~\bibnamefont{{Takiwaki}}},
  \bibinfo{author}{\bibfnamefont{S.~C.} \bibnamefont{{Whitehouse}}},
  \bibinfo{author}{\bibfnamefont{M.}~\bibnamefont{{Liebend{\"o}rfer}}},
  \bibnamefont{and} \bibinfo{author}{\bibfnamefont{K.}~\bibnamefont{{Sato}}},
  \bibinfo{journal}{PASJ} \textbf{\bibinfo{volume}{62}}, \bibinfo{pages}{L49}
  (\bibinfo{year}{2010}).

\bibitem[{\citenamefont{{Bruenn} et~al.}(2012)\citenamefont{{Bruenn},
  {Mezzacappa}, {Hix}, {Lentz}, {Bronson Messer}, {Lingerfelt}, {Blondin},
  {Endeve}, {Marronetti}, and {Yakunin}}}]{BrMeHi13}
\bibinfo{author}{\bibfnamefont{S.~W.} \bibnamefont{{Bruenn}}},
  \bibinfo{author}{\bibfnamefont{A.}~\bibnamefont{{Mezzacappa}}},
  \bibinfo{author}{\bibfnamefont{W.~R.} \bibnamefont{{Hix}}},
  \bibinfo{author}{\bibfnamefont{E.~J.} \bibnamefont{{Lentz}}},
  \bibinfo{author}{\bibfnamefont{O.~E.} \bibnamefont{{Bronson Messer}}},
  \bibinfo{author}{\bibfnamefont{E.~J.} \bibnamefont{{Lingerfelt}}},
  \bibinfo{author}{\bibfnamefont{J.~M.} \bibnamefont{{Blondin}}},
  \bibinfo{author}{\bibfnamefont{E.}~\bibnamefont{{Endeve}}},
  \bibinfo{author}{\bibfnamefont{P.}~\bibnamefont{{Marronetti}}},
  \bibnamefont{and} \bibinfo{author}{\bibfnamefont{K.~N.}
  \bibnamefont{{Yakunin}}}, \bibinfo{journal}{ApJ}
  \bibinfo{volume}{submitted} (\bibinfo{year}{2012}),
  \eprint{arXiv:1212.1747}.

\bibitem[{\citenamefont{{Woosley} and {Heger}}(2007)}]{WoHe07}
\bibinfo{author}{\bibfnamefont{S.~E.} \bibnamefont{{Woosley}}}
  \bibnamefont{and} \bibinfo{author}{\bibfnamefont{A.}~\bibnamefont{{Heger}}},
  \bibinfo{journal}{Phys. Rep.} \textbf{\bibinfo{volume}{442}},
  \bibinfo{pages}{269} (\bibinfo{year}{2007}).

\bibitem[{\citenamefont{Bruenn}(1985)}]{Brue85}
\bibinfo{author}{\bibfnamefont{S.~W.} \bibnamefont{Bruenn}},
  \bibinfo{journal}{ApJS} \textbf{\bibinfo{volume}{58}}, \bibinfo{pages}{771}
  (\bibinfo{year}{1985}).

\bibitem[{\citenamefont{Lattimer and Swesty}(1991)}]{LaSw91}
\bibinfo{author}{\bibfnamefont{J.}~\bibnamefont{Lattimer}} \bibnamefont{and}
  \bibinfo{author}{\bibfnamefont{F.~D.} \bibnamefont{Swesty}},
  \bibinfo{journal}{Nucl. Phys. A} \textbf{\bibinfo{volume}{535}},
  \bibinfo{pages}{331} (\bibinfo{year}{1991}).

\bibitem[{\citenamefont{{Cooperstein}}(1985)}]{Coop85}
\bibinfo{author}{\bibfnamefont{J.}~\bibnamefont{{Cooperstein}}},
  \bibinfo{journal}{Nucl. Phys. A} \textbf{\bibinfo{volume}{438}},
  \bibinfo{pages}{722} (\bibinfo{year}{1985}).

\bibitem[{\citenamefont{{Hix} and {Thielemann}}(1999)}]{HiTh99b}
\bibinfo{author}{\bibfnamefont{W.~R.} \bibnamefont{{Hix}}} \bibnamefont{and}
  \bibinfo{author}{\bibfnamefont{F.}~\bibnamefont{{Thielemann}}},
  \bibinfo{journal}{J. Comp. Appl. Math} \textbf{\bibinfo{volume}{109}},
  \bibinfo{pages}{321} (\bibinfo{year}{1999}).

\bibitem[{\citenamefont{{Buras} et~al.}(2006)\citenamefont{{Buras}, {Janka},
  {Rampp}, and {Kifonidis}}}]{BuJaRa06}
\bibinfo{author}{\bibfnamefont{R.}~\bibnamefont{{Buras}}},
  \bibinfo{author}{\bibfnamefont{H.-T.} \bibnamefont{{Janka}}},
  \bibinfo{author}{\bibfnamefont{M.}~\bibnamefont{{Rampp}}}, \bibnamefont{and}
  \bibinfo{author}{\bibfnamefont{K.}~\bibnamefont{{Kifonidis}}},
  \bibinfo{journal}{A\&A} \textbf{\bibinfo{volume}{457}}, \bibinfo{pages}{281}
  (\bibinfo{year}{2006}).

\bibitem[{\citenamefont{{Burrows} et~al.}(2007)\citenamefont{{Burrows},
  {Livne}, {Dessart}, {Ott}, and {Murphy}}}]{BuLiDe07}
\bibinfo{author}{\bibfnamefont{A.}~\bibnamefont{{Burrows}}},
  \bibinfo{author}{\bibfnamefont{E.}~\bibnamefont{{Livne}}},
  \bibinfo{author}{\bibfnamefont{L.}~\bibnamefont{{Dessart}}},
  \bibinfo{author}{\bibfnamefont{C.~D.} \bibnamefont{{Ott}}}, \bibnamefont{and}
  \bibinfo{author}{\bibfnamefont{J.}~\bibnamefont{{Murphy}}},
  \bibinfo{journal}{ApJ} \textbf{\bibinfo{volume}{655}}, \bibinfo{pages}{416}
  (\bibinfo{year}{2007}).

\bibitem[{\citenamefont{{Marek} and {Janka}}(2009)}]{MaJa09}
\bibinfo{author}{\bibfnamefont{A.}~\bibnamefont{{Marek}}} \bibnamefont{and}
  \bibinfo{author}{\bibfnamefont{H.-T.} \bibnamefont{{Janka}}},
  \bibinfo{journal}{ApJ} \textbf{\bibinfo{volume}{694}}, \bibinfo{pages}{664}
  (\bibinfo{year}{2009}).

\bibitem[{\citenamefont{{M{\"u}ller} et~al.}(2012)\citenamefont{{M{\"u}ller},
  {Janka}, and {Marek}}}]{MuJaMa12}
\bibinfo{author}{\bibfnamefont{B.}~\bibnamefont{{M{\"u}ller}}},
  \bibinfo{author}{\bibfnamefont{H.-T.} \bibnamefont{{Janka}}},
  \bibnamefont{and} \bibinfo{author}{\bibfnamefont{A.}~\bibnamefont{{Marek}}},
  \bibinfo{journal}{ApJ} \textbf{\bibinfo{volume}{756}}, \bibinfo{pages}{84}
  (\bibinfo{year}{2012}).

\bibitem[{\citenamefont{{Bruenn} et~al.}(2009)\citenamefont{{Bruenn},
  {Mezzacappa}, {Hix}, {Blondin}, {Marronetti}, {Messer}, {Dirk}, and
  {Yoshida}}}]{BrMeHi09b}
\bibinfo{author}{\bibfnamefont{S.~W.} \bibnamefont{{Bruenn}}},
  \bibinfo{author}{\bibfnamefont{A.}~\bibnamefont{{Mezzacappa}}},
  \bibinfo{author}{\bibfnamefont{W.~R.} \bibnamefont{{Hix}}},
  \bibinfo{author}{\bibfnamefont{J.~M.} \bibnamefont{{Blondin}}},
  \bibinfo{author}{\bibfnamefont{P.}~\bibnamefont{{Marronetti}}},
  \bibinfo{author}{\bibfnamefont{O.~E.~B.} \bibnamefont{{Messer}}},
  \bibinfo{author}{\bibfnamefont{C.~J.} \bibnamefont{{Dirk}}},
  \bibnamefont{and}
  \bibinfo{author}{\bibfnamefont{S.}~\bibnamefont{{Yoshida}}},
  \bibinfo{journal}{J. Phys.: Conf. Ser.} \textbf{\bibinfo{volume}{180}},
  \bibinfo{pages}{012018} (\bibinfo{year}{2009}).

\bibitem[{\citenamefont{{Burrows} et~al.}(2006)\citenamefont{{Burrows},
  {Livne}, {Dessart}, {Ott}, and {Murphy}}}]{BuLiDe06}
\bibinfo{author}{\bibfnamefont{A.}~\bibnamefont{{Burrows}}},
  \bibinfo{author}{\bibfnamefont{E.}~\bibnamefont{{Livne}}},
  \bibinfo{author}{\bibfnamefont{L.}~\bibnamefont{{Dessart}}},
  \bibinfo{author}{\bibfnamefont{C.~D.} \bibnamefont{{Ott}}}, \bibnamefont{and}
  \bibinfo{author}{\bibfnamefont{J.}~\bibnamefont{{Murphy}}},
  \bibinfo{journal}{ApJ} \textbf{\bibinfo{volume}{640}}, \bibinfo{pages}{878}
  (\bibinfo{year}{2006}), \eprint{astro-ph/0510687}.

\bibitem[{\citenamefont{{Ott} et~al.}(2008)\citenamefont{{Ott}, {Burrows},
  {Dessart}, and {Livne}}}]{OtBuDe08}
\bibinfo{author}{\bibfnamefont{C.~D.} \bibnamefont{{Ott}}},
  \bibinfo{author}{\bibfnamefont{A.}~\bibnamefont{{Burrows}}},
  \bibinfo{author}{\bibfnamefont{L.}~\bibnamefont{{Dessart}}},
  \bibnamefont{and} \bibinfo{author}{\bibfnamefont{E.}~\bibnamefont{{Livne}}},
  \bibinfo{journal}{ApJ} \textbf{\bibinfo{volume}{685}}, \bibinfo{pages}{1069}
  (\bibinfo{year}{2008}).

\bibitem[{\citenamefont{{Chertkow} et~al.}(2012)\citenamefont{{Chertkow},
  {Messer}, {Hix}, {Yakunin}, {Marronetti}, {Bruenn}, {Lentz}, {Blondin}, and
  {Mezzacappa}}}]{ChMeHi12}
\bibinfo{author}{\bibfnamefont{M.~A.} \bibnamefont{{Chertkow}}},
  \bibinfo{author}{\bibfnamefont{O.~E.~B.} \bibnamefont{{Messer}}},
  \bibinfo{author}{\bibfnamefont{W.~R.} \bibnamefont{{Hix}}},
  \bibinfo{author}{\bibfnamefont{K.}~\bibnamefont{{Yakunin}}},
  \bibinfo{author}{\bibfnamefont{P.}~\bibnamefont{{Marronetti}}},
  \bibinfo{author}{\bibfnamefont{S.~W.} \bibnamefont{{Bruenn}}},
  \bibinfo{author}{\bibfnamefont{E.~J.} \bibnamefont{{Lentz}}},
  \bibinfo{author}{\bibfnamefont{J.}~\bibnamefont{{Blondin}}},
  \bibnamefont{and}
  \bibinfo{author}{\bibfnamefont{A.}~\bibnamefont{{Mezzacappa}}},
  \bibinfo{journal}{J. Phys.: Conf. Ser.} \textbf{\bibinfo{volume}{402}},
  \bibinfo{pages}{012025} (\bibinfo{year}{2012}).

\bibitem[{\citenamefont{Kageyama and Sato}(2004)}]{KaSa04}
\bibinfo{author}{\bibfnamefont{A.}~\bibnamefont{Kageyama}} \bibnamefont{and}
  \bibinfo{author}{\bibfnamefont{T.}~\bibnamefont{Sato}},
  \bibinfo{journal}{Geochemistry, Geophysics, Geosystems}
  \textbf{\bibinfo{volume}{5}}, \bibinfo{pages}{Q09005} (\bibinfo{year}{2004}).

\end{thebibliography}


\end{document}